\documentclass[twocolumn,showpacs,preprintnumbers,amsmath,amssymb]{revtex4}
\usepackage{graphicx}% Include figure files
\usepackage{dcolumn}% Align table columns on decimal point
\usepackage{bm}% bold math

\def\be{\begin{equation}}
\def\ee{\end{equation}}
\def\bea{\begin{eqnarray}}
\def\eea{\end{eqnarray}}

\begin{document}

%\preprint{APS/123-QED}

\title{The twin paradox on the photon sphere}

\author{Marek A. Abramowicz}%%%%%%%%%%%%%%%%%%%%%%%%%%%%%%%
\email{ marek@fy.chalmers.se}
 \altaffiliation[Also at ]{Physics Department, G{\"o}teborg University, SE-412-96 G{\"o}teborg, Sweden}
\author{Stanis{\l}aw Bajtlik}%%%%%%%%%%%%%%%%%%%%%%%%%%%%%%
 \email{ bajtlik@camk.edu.pl}
\author{W{\l}odek Klu{\'z}niak}%%%%%%%%%%%%%%%%%%%%%%%%%%%%%%
 \email{ wlodek@camk.edu.pl}
\altaffiliation[Also at ]{Zielona G\'ora University, Wie\.za Braniborska, 
Lubuska 2, PL-65-265 Zielona G\'ora, Poland}
%\author{W{\l}odek Klu{\'z}niak}%%%%%%%%%%%%%%%%%%%%%%%%%%%%
% \email{ wlodek@camk.edu.pl}
%\affiliation[Also at ]{Zielona G\'ora University, Wie\.za Braniborska, Lubuska 2, %PL-65-265 Zielona G\'ora, Poland}

\affiliation{N. Copernicus Astronomical Center, Bartycka 18, 
PL-00-716 Warszawa, Poland, and Nordita, Copenhagen, Denmark}

\date{\today}

\begin{abstract}
We consider a new version of the twin paradox. The twins move along the
same circular free photon path around the Schwarzschild center. In this case,
despite their different velocities, all twins have the same non-zero
acceleration. On the circular photon path, the symmetry between the twins
situations  is broken not by acceleration (as it is in the case of the classic 
twin paradox), but by the existence of an absolute standard of rest (timelike
Killing vector). The twin with the higher velocity with respect to the standard 
of rest is younger on reunion. This closely resembles the case of periodic 
motions in compact (non-trivial topology) 3-D space recently considered in the
context of the twin paradox by Barrow and Levin \cite{barrow}, except that in
\cite{barrow} accelerations of all twins were equal to zero, and that in the 
case considered here, the 3-D space has trivial topology.
\end{abstract}

\pacs{04.20.Cv}% PACS, the Physics and Astronomy
\maketitle

The resolution of the classical twin paradox is that the travelling twin must
suffer an acceleration since he stops, turns around and returns to his never
accelerating twin brother. The physical situation of the twins is therefore not
the same, and indeed, the one who accelerates is younger on reunion. Barrow 
and Levin \cite{barrow}, see also \cite{bajtlik}, discussed a clever version 
of the paradox in compact spaces, with  the twins moving with zero acceleration 
on a closed path. Compactness of the space defines a global standard of rest, 
and the twin moving faster with respect to this standard is younger on reunion. 
In this Note we consider still another example of the twin paradox. Twins move 
with constant velocity along a circular photon path in the Schwarzschild spacetime. 
In this case, as shown by Abramowicz and Lasota \cite{lasota}, see also
\cite{carter}, they all have the same non-zero acceleration. We will show that 
the one who moves faster with respect to the standard of rest (given by the 
Killing symmetry of the Schwarzschild spacetime) is younger on reunion. 

In Schwarzschild coordinates, the Schwarzschild metric has the form,
\bea %%%%%%%%%%%%%%%%%%%%%%%%%%%%%%%%%%%%%%%%%%%%%%%%%%%%%%
ds^2 &=& \left( 1 - \frac{r_G}{r} \right)\,dt^2 \nonumber \\ 
&-& \left( 1 - \frac{r_G}{r}
\right)^{-1}\,dr^2 - r^2\,\left( d\theta^2 + \sin^2 \theta \, 
d\varphi^2 \right).
%\nonumber
\eea %%%%%%%%%%%%%%%%%%%%%%%%%%%%%%%%%%%%%%%%%%%%%%%%%%%%%%%
The metric depends neither on the time coordinate $t$, nor on the azimuthal
coordinate $\varphi$. It is convenient to invariantly express these time and
azimuthal symmetries in terms of two Killing vector fields,
\be %%%%%%%%%%%%%%%%%%%%%%%%%%%%%%%%%%%%%%%%%%%%%%%%%%%%%%%
\eta^i = \delta^i_{~t}, ~~~\xi^i = \delta^i_{~\varphi}, ~~~\eta^i\xi_i = 0.
\ee %%%%%%%%%%%%%%%%%%%%%%%%%%%%%%%%%%%%%%%%%%%%%%%%%%%%%%%
Note that $\eta^i\eta_i = g_{tt} = 1-r_G/r$, $\xi^i\xi_i = g_{\varphi\varphi}
= - \sin^2 \theta \,r^2$. The unit timelike vector,
\be %%%%%%%%%%%%%%%%%%%%%%%%%%%%%%%%%%%%%%%%%%%%%%%%%%%%%%%
N^i = (\eta^k\eta_k)^{-1/2}\, \eta^i,
\ee %%%%%%%%%%%%%%%%%%%%%%%%%%%%%%%%%%%%%%%%%%%%%%%%%%%%%%%
defines the static observers outside the event horizon $r=r_G$. They
provide the absolute standard of rest in Schwarzschild spacetime.

Circular orbits $r =$const on equatorial plane $\theta = \pi/2$ have the only 
non zero components $U^t = dt/ds \equiv A$ and $U^{\varphi} = d\varphi/ds \equiv
A\Omega$. They are invariantly defined by,
\be %%%%%%%%%%%%%%%%%%%%%%%%%%%%%%%%%%%%%%%%%%%%%%%%%%%%%%%
\label{velocity}
U^i = A\,\left( \eta^i + \Omega \xi^i \right)
\ee %%%%%%%%%%%%%%%%%%%%%%%%%%%%%%%%%%%%%%%%%%%%%%%%%%%%%%%
Since $U^iU_i=1$, the normalization is $1/A^2 = \eta^i\eta_i + 
\Omega^2 \xi^i\xi_i$.

The velocity with respect to the absolute standard of rest 
(static observers) is defined by,
\be %%%%%%%%%%%%%%%%%%%%%%%%%%%%%%%%%%%%%%%%%%%%%%%%%%%%%%%
V^i = \left( \delta^i_{~k} - N^iN_k\right)\,U^k, 
\ee %%%%%%%%%%%%%%%%%%%%%%%%%%%%%%%%%%%%%%%%%%%%%%%%%%%%%%%
and the corresponding speed $V$ (or ${\tilde V}$) is defined by
\be %%%%%%%%%%%%%%%%%%%%%%%%%%%%%%%%%%%%%%%%%%%%%%%%%%%%%%%
{\tilde V}^2 = \frac {V^2}{1-V^2} = -V^iV_i.
\ee %%%%%%%%%%%%%%%%%%%%%%%%%%%%%%%%%%%%%%%%%%%%%%%%%%%%%%%
Note that $V$$=$$c\beta$ and ${\tilde V}$$=$$c\beta\gamma$ in terms of
the Lorentz $\beta$ and $\gamma$$\equiv$$(1-\beta^2)^{-1/2}$.
From the above definitions one calculates,
\be %%%%%%%%%%%%%%%%%%%%%%%%%%%%%%%%%%%%%%%%%%%%%%%%%%%%%%%
V^2 = - \Omega^2 \frac {\xi^i\xi_i}{\eta^k\eta_k} \equiv
\Omega^2 \,{\cal R}^2 = \Omega^2\,\frac {r^3}{r-r_G}.
\ee %%%%%%%%%%%%%%%%%%%%%%%%%%%%%%%%%%%%%%%%%%%%%%%%%%%%%%%
The quantity ${\cal R}$ is the ``distance from the rotation 
axis''.

The acceleration on an orbit described by (\ref{velocity}) is
\be %%%%%%%%%%%%%%%%%%%%%%%%%%%%%%%%%%%%%%%%%%%%%%%%%%%%%%%
\label{acceleration}
a_i \equiv U^k\,\nabla_k\,U_i = -\frac{1}{2}\, \frac
{\nabla_i (\eta^k \eta_k) + \Omega^2\,\nabla_i (\xi^k\xi_k)}
{(\eta^k \eta_k) + \Omega^2\,(\xi^k\xi_k)}.
\ee %%%%%%%%%%%%%%%%%%%%%%%%%%%%%%%%%%%%%%%%%%%%%%%%%%%%%%%
One may define the ``gravitational potential'' $\Phi$ by the 
invariant expression
\be %%%%%%%%%%%%%%%%%%%%%%%%%%%%%%%%%%%%%%%%%%%%%%%%%%%%%%%
\Phi = -\frac{1}{2}\,\ln (\eta^k\eta_k),
\ee %%%%%%%%%%%%%%%%%%%%%%%%%%%%%%%%%%%%%%%%%%%%%%%%%%%%%%%
and then put the acceleration formula (\ref{acceleration}) 
into a form identical with its Newtonian version,
\be %%%%%%%%%%%%%%%%%%%%%%%%%%%%%%%%%%%%%%%%%%%%%%%%%%%%%%%
\label{acceleration-Newtonian}
a_i = 
\nabla_i \Phi + \frac{{\tilde V}^2}{{\cal R}}\nabla_i\,{\cal R}.
\ee %%%%%%%%%%%%%%%%%%%%%%%%%%%%%%%%%%%%%%%%%%%%%%%%%%%%%%%
From this equation one calculates that for the free (geodesic)
motion, the orbital velocity $V^2$ equals in the Schwarzschild 
spacetime,
\be %%%%%%%%%%%%%%%%%%%%%%%%%%%%%%%%%%%%%%%%%%%%%%%%%%%%%%%
V^2 = \frac{1}{2}\frac{r_G}{r - r_G}.
\ee %%%%%%%%%%%%%%%%%%%%%%%%%%%%%%%%%%%%%%%%%%%%%%%%%%%%%%%
In particular, for $r = (3/2)r_G$ it is $V^2 = 1$, which
therefore corresponds to a free, circular motion of photons.
The $r = (3/2)r_G$ circles are called the photon circles,
and one also speaks about the photon sphere.

Abramowicz and Lasota \cite{lasota} noticed that $\nabla_i\,{\cal R}
= 0$ on the photon circle, and they deduced from 
(\ref{acceleration-Newtonian}) that on the photon circle 
acceleration does not depend on the orbital velocity:
for all steady, non-geodesic, circular motions along 
$r = (3/2)r_G$ the acceleration takes the universal value 
(we give its $r$ component only, as all other components 
are zero),
\be %%%%%%%%%%%%%%%%%%%%%%%%%%%%%%%%%%%%%%%%%%%%%%%%%%%%%%%
\label{universal}
\left(a_r\right)^{*} = 
-\frac{1}{2}\,\left[ \frac{d\ln (1 - r_G/r)}{dr}\right]_{(3/2)r_G} 
= ~-\frac{\,2}{3\,r_G}.
\ee %%%%%%%%%%%%%%%%%%%%%%%%%%%%%%%%%%%%%%%%%%%%%%%%%%%%%%% 
This suggests another version of the twin paradox. At the time $t$$=$$t_0$
two identical twins meet. Because they move with two different, constant, 
orbital velocities $V_1$ and $V_2$ along the same photon circle, they will
certainly meet again. Who will be younger on reunion at the time 
$t=t_0 + \Delta\,t$? Unlike in the classical version of the twin paradox,
but similarly as in the case of compact space considered by Barrow 
and Levin \cite{barrow}, the acceleration $a$ cannot be the answer here, as the two twins 
have {\it the same acceleration}. In the compact space they all have $a$$=$$0$, 
and on the photon circle they have $a$$=$$a^{*}$ given by equation
(\ref{universal}). Indeed, like in the compact space, the answer is the velocity,
the {\it absolute} velocity with respect to the {\it absolute} standard of rest:
as we shall see, on reunion the twin who moves faster with
respect to the standard of rest is younger. In the compact space, as explained 
by Barrow and Levin, there is no local standard of rest, but there is a global 
one. In the case of photon circle, the standard of rest is local, and given by 
the Killing symmetry.

The proper time elapsed from separation to reunion is,
\be %%%%%%%%%%%%%%%%%%%%%%%%%%%%%%%%%%%%%%%%%%%%%%%%%%%%%%%
\tau = \int_{t_0}^{t_0+\Delta t}\,\frac{ds}{dt}\,dt = 
\int_{t_0}^{t_0+\Delta t}\,\frac{dt}{U^t} =
\frac{\Delta t}{g_{tt}^{1/2}\gamma}.
\ee %%%%%%%%%%%%%%%%%%%%%%%%%%%%%%%%%%%%%%%%%%%%%%%%%%%%%%%
From this, one finally deduces,
\be %%%%%%%%%%%%%%%%%%%%%%%%%%%%%%%%%%%%%%%%%%%%%%%%%%%%%%%
\tau_1 = \frac{\gamma_2}{\gamma_1}\tau_2,
\ee %%%%%%%%%%%%%%%%%%%%%%%%%%%%%%%%%%%%%%%%%%%%%%%%%%%%%%%
which proves the point: the twin who moves faster is younger
on reunion.

\begin{acknowledgments}
We acknowledge the support from Polish State Committee for Scientific
Research, through grants 1P03D~012~26, 1PD3D~005~30, and N203~009~31/1466.
\end{acknowledgments}

\bibliography{twin}

\end{document}